\documentstyle[12pt,epsf,psfig]{article}
\def\J{$J/\psi$}
\def\j{J/\psi}
\def\X{$\chi$}
\def\x{\chi}
\def\P{$\psi'$}
\def\p{\psi'}

\def\C{c{\bar c}}

\def\e{\epsilon}
\def\L{\Lambda_{\rm QCD}}

\def\be{\begin{equation}}
\def\ee{\end{equation}}

\def\lsim{\raise0.3ex\hbox{$<$\kern-0.75em\raise-1.1ex\hbox{$\sim$}}}
\def\gsim{\raise0.3ex\hbox{$>$\kern-0.75em\raise-1.1ex\hbox{$\sim$}}}

\def\NP{{ Nucl.\ Phys.\ }}
\def\PL{{ Phys.\ Lett.\ }}
\def\PR{{ Phys.\ Rev.\ }}

\def\PRL{{ Phys.\ Rev.\ Lett.\ }}

\def\ZP{{ Z.\ Phys.\ }}

\begin{document}
\noindent November 1997 \hfill BI-TP 97/47 

\vskip 1.5cm

\centerline{\large{\bf Colour Deconfinement and \J~Suppression}}

\medskip

\centerline{\large{\bf in High Energy Nuclear Collisions}$^*$}

\vskip 1cm

\centerline{\bf Helmut Satz}

\bigskip

\centerline{Fakult\"at f\"ur Physik, Universit\"at Bielefeld}

\par

\centerline{D-33501 Bielefeld, Germany}

\vskip 1.5cm
\leftskip 2.0cm

\centerline{\bf Contents:}

\bigskip

1.\ Introduction

\medskip

2.\ Charmomium Dissociation and Colour Deconfinement

\medskip

3.\ \J~Production in Nuclear Collisions

\medskip

4.\ Anomalous \J~Suppression

\medskip

5.\ Outlook and Summary

\leftskip 0cm

\vfill

\noindent
{\bf 1.\ Introduction}

\medskip

Strong interaction thermodynamics deals with the behaviour of matter at
extreme temperatures and densities. Its central theme is the transition
from hadronic matter to a new state, the quark-gluon plasma, in which
quarks are no longer confined to colour-neutral bound states. The
existence of the quark-gluon plasma (QGP) is one of the basic predictions
of quantum chromodynamics (QCD), and its experimental observation
represents one of the great challenges to present high energy physics.

\medskip

Colour deconfinement is essentially the QCD version of the
insulator-conductor transition familiar from condensed matter physics:
at sufficiently high colour charge densities, screening suppresses the
long-range confining part of the strong interaction,
\be
\sigma r ~\to ~\sigma r \left[ {1 - e^{-\mu r} \over \mu r} \right],
\label{1}
\ee
where $\mu^{-1}$ is the screening radius. Such screening leads to the
melting of hadronic states, as illustrated in Fig.\ 1.

\bigskip

\hrule

~\par

\noindent
*) Lecture given at the 35th Course of the {\sl International School of
Subnuclear Physics}, Erice/Sicily, 26.\ 8.\ -- 4.\ 9.\ 1997, to appear
in the Proceedings.

\newpage

\bigskip

\begin{figure}[h]
\vspace*{-0mm}
\centerline{\psfig{file=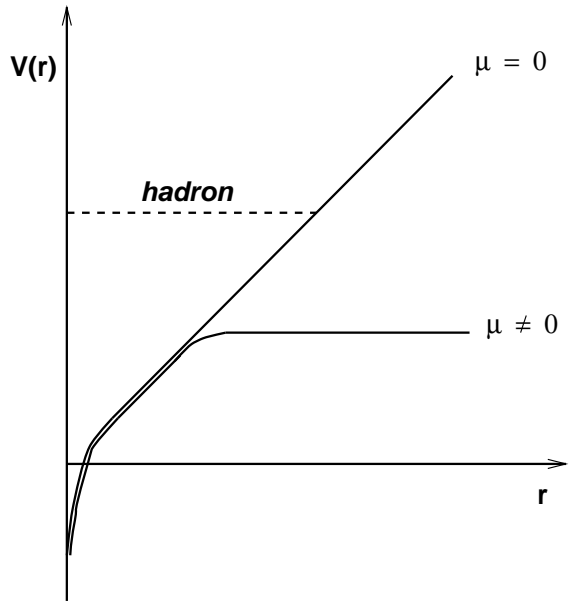,height= 80mm,angle= -90}}
\caption{Deconfinement by colour screening.}
\end{figure}

\medskip

The computer simulation of finite temperature QCD on the lattice allows
an {\sl ab initio} study of this phenomenon, starting from the QCD
Lagrangian as dynamical basis. We cite here only the most important
results of such studies \cite{LQCD}. As seen in Fig.\ 2, the energy
density of strongly interacting matter shows a sudden increase at a
temperature $T_c \simeq 150$ MeV; from a value near that for an ideal
gas of pions, it jumps to that for an ideal plasma of quarks and
gluons \cite{MILC}. The associated pressure, however, follows this
increase more slowly; since $\e = 3 P$ for an ideal gas of massless
constituents, this indicates that at least until $T/T_c \simeq 2 - 3$
there must still be considerable remnant interactions in the plasma. In
Fig.\ 3, we see that the increase of the
energy density is indeed due to deconfinement. The order parameter for
quark binding is $ \langle L \rangle \sim \exp-\{V(\infty)/T\}$, where $V(\infty)$ is
the potential of a static quark-antiquark pair in the limit of infinite
separation. In confined matter, $V(\infty)$ diverges, causing $ \langle L \rangle$ to
vanish; in a QGP, screening keeps $V(\infty)$ and hence also $ \langle L
\rangle$ finite. We see that the transition for $ \langle L \rangle$
coincides with the increase of $\e$. Moreover, we note that at the same
temperature the effective quark mass, measured by $\langle \psi {\bar
\psi} \rangle$, drops from the finite value associated with constituent
quarks to the essentially vanishing value of the light $u$ and $d$
quarks. From this, we can conclude that in the regime studied here
(vanishing overall baryon number density), deconfinement and chiral
symmetry restoration coincide.

\begin{figure}[p]
\vspace*{-0mm}
\centerline{\psfig{file=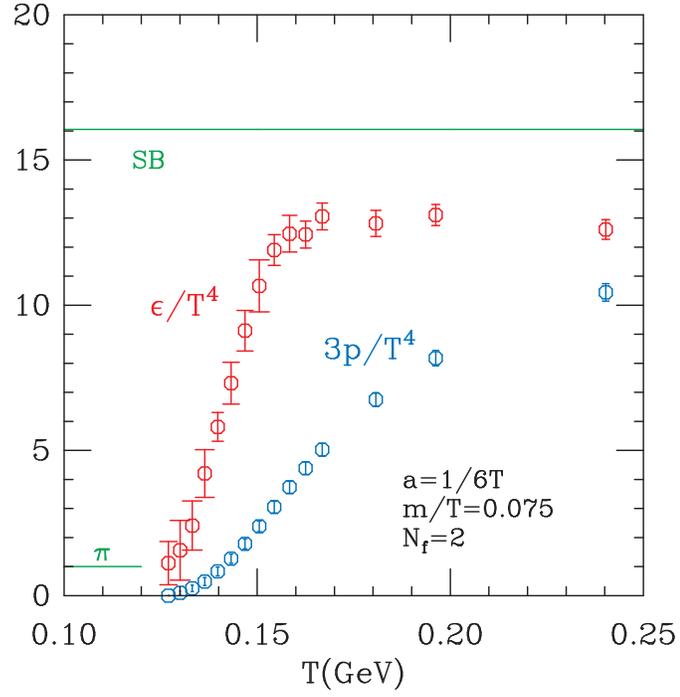,height= 90mm}}
\caption{ Energy density and pressure in two-flavour
QCD \cite{MILC}.}
\end{figure}

\begin{figure}[p]
\vspace*{-0mm}
\centerline{\psfig{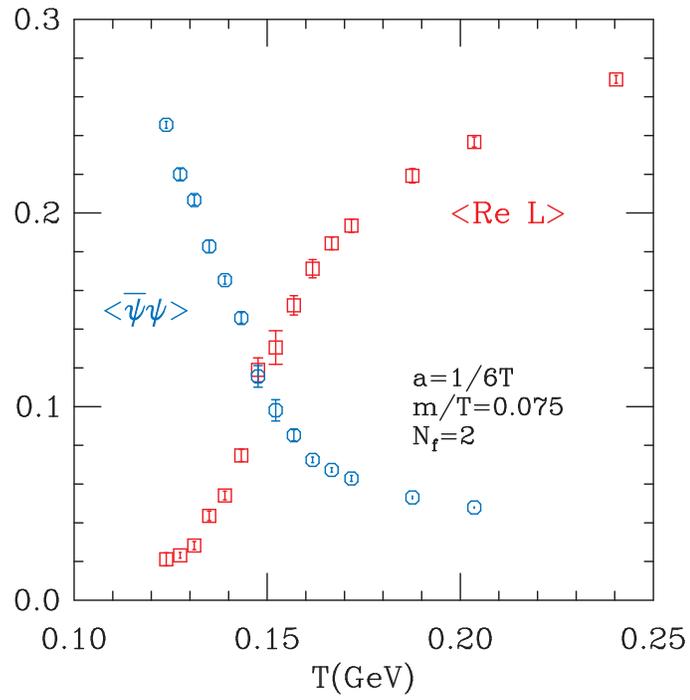}}
\caption{Deconfinement and chiral symmetry in
two-flavour QCD \cite{MILC}.}
\end{figure}

\medskip

Lattice QCD thus predicts that the transition from hadronic matter to a
QGP occurs at a temperature of around 150 - 200 MeV. The uncertainty
in this value is due to finite lattice sizes and to uncertainties in
the calculation of hadron masses; both can be removed in the next few
years by improved computer technology. In any case, we know already now
(see Fig.\ 2) that in order to produce a quark-gluon plasma, energy
densities of about 1 - 3 GeV/fm$^3$ are necessary. Shortly after the
big bang, our universe must have consisted of such matter, and it would
certainly be interesting to reproduce it in the laboratory.

\newpage

High energy nuclear collisions are expected to provide the required
conditions. When two nuclei collide at high energy, they pass through
each other and leave behind a trail of locally disturbed vacuum, of energy
deposited in bubbles of strongly interacting matter; these bubbles are
our candidates for the `little bang'. They were formed shortly after
the passage of two nuclei, and the first question is whether they were
sufficiently hot and dense to reach colour deconfinement. In any case,
since they are not contained by anything, they will expand, cool off and
eventually freeze out into free hadrons, which can be observed
experimentally. We can use these hadrons to estimate the initial energy
density of the bubbles \cite{Bj}, by letting the evolution film run
backwards. The observed number $(dN/dy)$ of produced hadrons, each of
average energy $p_0$,
must have originated in a volume of transverse size determined by the
colliding nuclei; the longitudinal size can be estimated if the bubbles
are defined to give isotropic momentum distributions. In this way, we 
obtain for an $A-A$ collision
\be
\e_0 \simeq \left( {dN \over dy} \right)_{y=0} {p_0 \over \pi R_A^2
\tau_0}, \label{2}
\ee
where $R_A\simeq 1.12 A^{1/3}$ and $\tau_0 \simeq 1$ fm. Using the
multiplicites measured at CERN-SPS energies, we get energy densities in
the range 2 - 5 GeV/fm$^3$; estimates for RHIC and LHC run as high as 10
- 20 GeV/fm$^3$. Thus there seems to be a real chance to produce and
studythe QGP in the laboratory, making the quark-hadron transition
the only experimentally accessible cosmological phase transition.

\medskip

The fundamental question facing us is therefore how to check if the
medium produced in high energy nuclear collisions was deconfined in its
early stages. To show how non-trivial this problem is, we recall that no
measurement on a glass of water can ever tell us if it was ice half an
hour ago. We want to find out what things were like {\sl before} the
time at which we can carry out measurements. A useful deconfinement
probe thus has to fulfill a number of conditions:
\begin{itemize}
\item{It must be present early and retain its memory throughout the
evolution;}
\item{it must be hard enough to resolve the short-distance scales
of the QGP; and}
\item{it must be able to distinguish confined and deconfined media.}
\end{itemize}
\noindent
So far, two candidates have been proposed to do this:
\begin{itemize}
\item{quarkonium states, whose dissociation pattern is quite different
in confined and deconfined media (``\J~suppression"\cite{Matsui}), and}
\item{hard jets, whose energy loss and transverse momentum broadening
depends on the nature of the medium they traverse
(``jet quenching"\cite{Bj-jet}).}
\end{itemize}
\noindent
We shall here consider the first of these two and study quarkonium
dissociation in nuclear collisions as test for colour deconfinement. It
gives rather clear-cut theoretical predictions for experiments
accessible at SPS energies; jets will probably have to wait for RHIC or
LHC studies.

\medskip

Before turning to the details of the probe, we restate the basic
question: Given a box of unidentified matter, we want to
know if the quarks and gluons which make up this matter are in a
confined or deconfined state. It is thus not evidence for
quarks and gluons that we look for, but rather the
confinement/deconfinement status of these elementary building blocks
of all forms of matter.

\bigskip

\noindent
{\bf 2.\ Charmonium Dissociation and Colour Deconfinement}

\bigskip

As prototype for charmonium, consider the \J; it is the $1S$ bound state
of a charm quark ($m_c \simeq 1.4$ GeV) and its antiquark, with $M_{\j}
\simeq 3.1$ GeV. Its usefulness as deconfinement probe is easily seen.
If a \J~is placed into a hot medium of deconfined quarks and gluons,
colour screening will
dissolve the binding, so that the $c$ and the $\bar c$ separate. When
the medium cools down to the confinement transition point, they will
therefore in general be too far apart to see each other. Since thermal
production of further $\C$ pairs is negligibly small because of the
high charm quark mass, the $c$ must combine with a light antiquark to
form a $D$, and the $\bar c$ with a light quark for a $\bar D$. The
presence of a quark-gluon plasma will thus lead to a suppression of
\J~production \cite{Matsui}.

\medskip

We shall now first consider this \J~suppression in terms of colour
screening, i.e., as consequence of global features of the medium, and
then turn to a microscopic approach, in which the bound state is
assumed to be broken up by collisions with constituents of the medium.

\medskip

\noindent{\bf 2a.\ Colour Screening}

\medskip

Because of the large charm quark mass, the charmonium spectrum can be
calculated with good precision by means of the Schr\"odinger equation
\cite{Schroedinger}
\be
[2m_c + {1 \over m_c}\nabla^2 + V(r)] \Psi_{n,l} = M_{n,l} \Psi_{n,l},
\label{3}
\ee
where the potential $V(r)=\sigma r - \alpha/r$ contains a confining
long-distance part $\sigma r$ and a Coulomb-like short-distance term
$\alpha/r$. For different values of the principal quantum number $n$
and the orbital quantum number $l$, the masses $M_{n,l}$ and the wave
functions $\Psi_{n,l}(r)$ of different charmonium states
$\j,~\chi, ~\psi',~...$ in vacuum are given in terms of the
constants $m_c,~\sigma$ and $\alpha$.

\medskip

In a medium, the potential becomes screened,
\be
V(r,\mu) = {\sigma \over \mu}[1 - e^{-\mu r}]  - {\alpha \over r}
e^{-\mu r}, \label{4}
\ee
where $\mu$ is the screening mass, i.e., $r_D = \mu^{-1}$ is the `Debye'
colour screening radius. Screening is a global feature of the medium,
shortening the range of the binding potential. Once $\mu$ becomes
sufficiently large,
the bound states begin to disappear, starting with the most weakly
bound; hence for $\mu \geq \mu_d^i$, the bound state $i$ is no longer
possible \cite{KMS}.

\medskip

Using finite temperature lattice QCD, we can now determine the screening
mass $\mu(T)$ as function of the temperature $T$ or, equivalently, as
function of the energy density $\e$ of the medium \cite{LGT-mu}. The
resulting melting pattern for the most important charmonium states is
summarized in Table 1; we see that while both \P~and \X~melt at the
critical deconfinement point, the \J, being smaller and more tightly
bound, survives to about 1.2$T_c$ and hence about twice the critical
energy density. With increasing temperature, a hot medium will thus lead
to successive charmonium melting, so that the
suppression or survival of specific charmomium states serves as a
thermometer for the medium, in much the same way as the relative
intensity of spectral lines in stellar interiors measure the temperature
of stellar matter \cite{Kajantie}. Note, however, that other possible
sources of charmonium dissociation have to be considered before the
method becomes unambiguous. This will be done in the next subsection.

\medskip

\begin{center}
\begin{tabular}{|c||c|c|c|}
\hline
&  &  & \\
 state & $\mu_d$~[{\rm GeV}] & $T_d$ & $\e_d$ \\
&  &  & \\
\hline
\hline
\J & 0.70 & 1.2~$T_c$ & 2 $\e_c$ \\
\hline
\X & 0.35 & $T_c$     & $\e_c$   \\
\hline
\P & 0.35 & $T_c$     & $\e_c$   \\
\hline
\end{tabular}\end{center}

\centerline{Table 1: Charmonium Dissociation by Colour Screening}

\bigskip

\medskip

\par

\noindent
{\bf 2b.\ Gluon Dissociation}

\medskip

The binding energy of the \J, i.e., the energy difference between the
\J~mass and the open charm threshold, is with
$\Delta E_{\j}=2M_D - M_{\j} \simeq 0.64~{\rm GeV} \gg \L$
considerably larger than the typical non-perturbative hadronic scale
$\L \simeq 0.2$ GeV. As a consequence, the size of the \J~is much
smaller than that of a typical hadron, $r_{\j} \simeq 0.2~{\rm fm} \ll
\L^{-1} = 1~{\rm fm}$. Hence the \J~is a hadron with characteristic
short-distance features; in particular, rather hard gluons are
necessary to resolve or dissociate it, making such a dissociation
accessible to perturbative calculations. \J~collisions with ordinary
hadrons made up of the usual $u,d$ and $s$ quarks thus probe the local
partonic structure of these `light' hadrons, not their global hadronic
aspects, such as mass or size. It is for this reason that \J's can be
used as confinement/deconfinement probe.

\medskip

This can be illustrated by a simple example. Consider an ideal pion gas
as a confined medium. The momentum spectrum of pions has the Boltzmann
form $f(p) \sim \exp-(|p|/T)$, giving the pions an average momentum
$\langle |p| \rangle = 3~T$. With the pionic gluon distribution function
$xg(x) \sim (1-x)^3$, where $x=k/p$ denotes the fraction of the pion
momentum carried by a gluon, the average momenta of gluons confined to
pions becomes
\be
\langle |k| \rangle_{\rm conf}  \simeq 0.6~T. \label{5}
\ee
On the other hand, an ideal QGP as prototype of a deconfined medium
gives the gluons themselves the Boltzmann distribution $f(k) \sim
\exp-(|k|/T)$ and hence average momenta
\be
\langle |k| \rangle_{\rm deconf} = 3~T. \label{6}
\ee
Deconfinement thus results in a hardening of the gluon momentum
distribution. More generally speaking, the onset of deconfinement will
lead to parton distribution functions which are different from those
for free hadrons, 
as determined by deep inelastic scattering experiments.
Since hard gluons are needed to resolve and dissociate \J's, one can use
\J's to probe the in-medium gluon hardness and hence the confinement
status of the medium.

\medskip

These qualitative considerations can be put on a solid theoretical basis
provided by short-distance QCD \cite{Peskin} -- \cite{KS3}. In Fig.\ 4
we show the relevant diagram for the calculation of the inelastic
\J-hadron cross section, as obtained in the operator product expansion
(essentially a multipole expansion for the charmonium quark-antiquark
system). The upper part of the figure shows \J~dissociation by gluon
interaction; the cross section for this process,
\be
\sigma_{g-\j} \sim (k-\Delta E_{\j})^{3/2}  k^{-5}, \label{7}
\ee
constitutes the QCD analogue of the photo-effect. Convoluting the \J~
gluon-dissociation with the gluon distribution in the incident hadron,
$xg(x) \simeq 0.5(1-x)^n$, we obtain
\be
\sigma_{h-\j} \simeq \sigma_{\rm geom} (1 - \lambda_0/\lambda)^{n+2.5}
\label{8}
\ee
for the inelastic \J-hadron cross section, with $\lambda \simeq
(s-M_{\psi}^2)/M_{\psi}$ and $\lambda_0 \simeq (M_h + \Delta E_{\psi}$);
$s$ denotes the squared \J-hadron collision energy. In Eq.\ (8),
$\sigma_{\rm geom} \simeq {\rm const}.\ r_{\j}^2 \simeq 2 - 3$ mb is
the geometric cross section obtained with the
mentioned gluon distribution. In the threshold region and for
relatively low collision energies, $\sigma_{h-\j}$ is very strongly
damped because of the suppression $(1-x)^n$ of hard gluons in hadrons,
which leads to the factor $(1 - \lambda_0/\lambda)^{n+2.5}$ in Eq.\ (8).
In Fig.\ 5, we compare the cross sections for \J~dissociation by gluons
(``gluo-effect") and by pions ($n=3$), as given by Eq's (7) and (8).
Gluon dissociation shows the typical photo-effect form, vanishing until
the gluon momentum $k$ passes the binding energy $\Delta E_{\psi}$;
it peaks just a little later and then vanishes again when
sufficiently hard gluons just pass through the much larger charmonium
bound states. In contrast, the \J-hadron cross section remains
negligibly small until rather high hadron momenta (3 - 4 GeV). In a
thermal medium, such momenta correspond to temperatures of more than one
GeV. Hence confined media in the temperature range of a few hundred MeV
are essentially transparent to \J's, while deconfined media of the
same temperatures very effectively dissociate them and thus are
\J-opaque.

\bigskip

\begin{figure}[h]
\vspace*{-0mm}
\centerline{\psfig{file=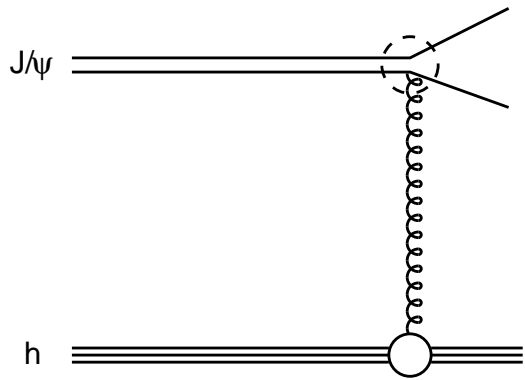,height=50mm}}
\caption{\J~dissociation by hadron interaction.}
\end{figure}

\medskip

The situation for \X's is quite similar, except for an earlier gluon
dissociation threshold due to the lower \X~binding energy of about 250
MeV. This difference provides the micro\-scopic basis for the successive
melting of different charmonium states noted above. -- For the \P,
however, we have an almost negligible binding energy of only 60 MeV,
so that there cannot really be any difference in its dissociation by
confined or deconfined media. In other words, any strongly interacting
matter is expected to be \P-opaque.

\medskip

\begin{figure}[h]
\vspace*{-0mm}
\centerline{\psfig{file=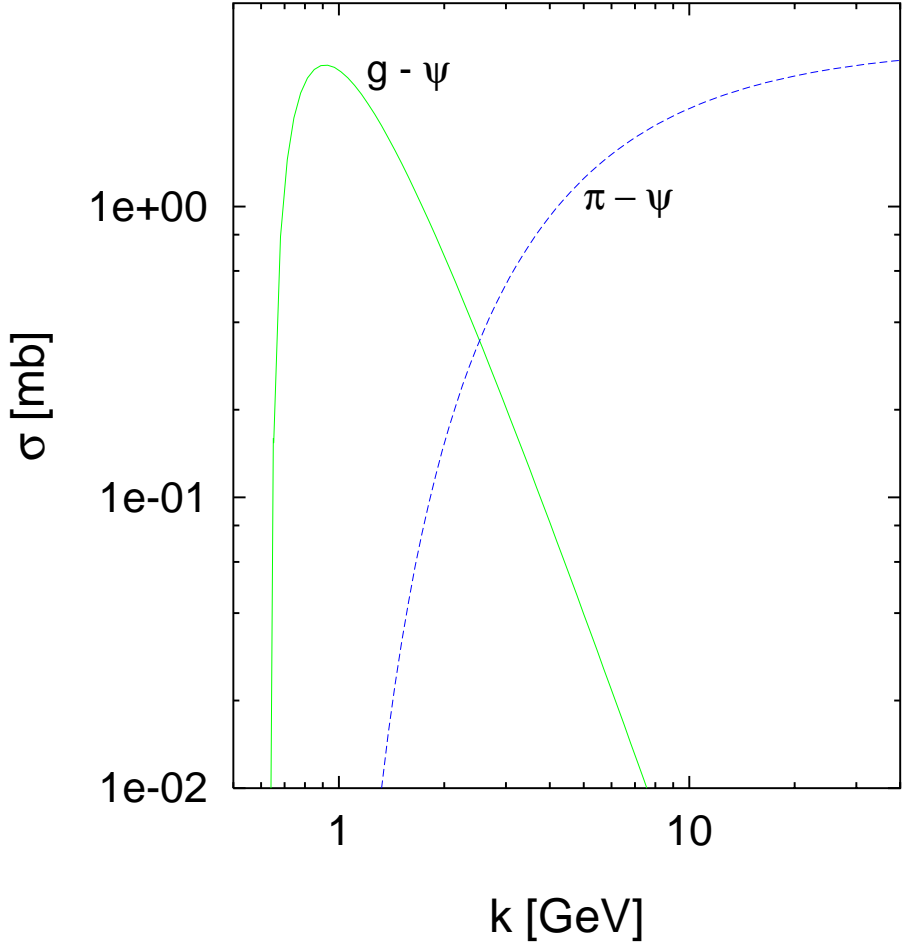,height= 130mm,angle= -90}}
\caption{\J~dissociation by gluons and by pions; $k$ denotes the
momentum of the projectile incident on a stationary \J.}
\end{figure}

\medskip

We can thus define a schematic test to determine the confinement status
of a box of unidentified matter. We first shine a \P~beam at it: if it
is transparent to this, the box does not contain strongly interacting
matter; otherwise it does. In that case we repeat the procedure with a
\J~beam: if this is unaffected, the medium in the box is confined; if
the beam is attenuated, it is deconfined. The problem of an unambiguous
deconfinement test is thus in principle solved: there is \J~dissociation
if and only if the medium is deconfined. However, the test pre-supposes
the existence of a prepared strongly interacting medium and the
availability of \J~and \P~beams as probes, and in nuclear collisions,
neither of these is immediately given.

\bigskip

\noindent
{\bf 3.\ \J~Production in Nuclear Collisions}

\medskip

In nuclear collisions, both the charmonium states and the matter to be
probed are produced in the course of the collision, and both require
a finite `formation time' to be formed. We therefore have to ask what
will happen `before' they are there and consider in particular
pre-resonance absorption in normal nuclear matter.

\medskip

Quarkonium production in hadron-hadron collisions has in recent years
been studied quite extensively, triggered in particular by detailed and
quite conclusive experiments at FNAL \cite{FNAL}. The production of a
\J~in a $p-p$ or $p-{\bar p}$ collision begins with the production of a
$\C$ pair,
which occurs at high energies dominantly by gluon fusion (Fig.\ 6). The
$\C$ is generally in a colour octet state and has to neutralize its
colour in order to leave the interaction region and form a \J. The
colour singlet model \cite{singlet} proposed a perturbative
treatment of this colour neutralisation; this is now clearly ruled out
by the mentioned FNAL experiments. Colour neutralisation thus takes
place in a non-perturbative way \cite{evaporation}, and the recently
proposed colour octet formalism \cite{octet} can be extended to obtain 
such a description \cite{KS6}. 
The coloured $\C$ binds with a soft
collinear gluon to form a colour singlet $\C-g$ state; in a proper
time $\tau_{\C g} \simeq (2m_c \L)^{-1/2} \simeq 0.3$ fm, this falls
into the
$\C$ singlet state which is the basic component of a physical \J.
Since the life-time (and hence also the size) of this
pre-resonance $\C g$ state is determined by the hadronic scale $\L$, it
is the same for \J, \X~and \P; in contrast, the final resonances
have very different geometric sizes.

\bigskip

\begin{figure}[h]
\vspace*{-0mm}
\epsfysize=7cm
\centerline{\epsffile{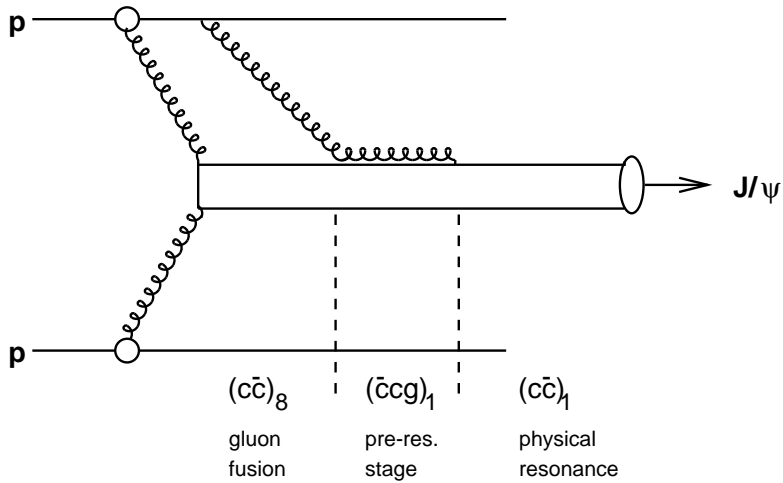}}
\caption{\J~production by $p-p$ collision.}
\end{figure}

\medskip

This formation process has rather striking consequences for charmonium
production in $p-A$ collisions. The presence of the nuclear target
medium is known to reduce \J~production rates in $p-A$ collisions
\cite{NA3,E772}, relative to those in $p-p$ interactions (Fig.\ 7).
However, these experiments are
carried out in a kinematic region giving the nascent \J~momenta of 50
GeV/c or more in the target rest frame. As a result, the transition $\C
g \to \j,~\x$ or \P~occurs outside the target nucleus; the nuclear
matter
of the target sees only the passage of the $\C g$ state. Hence the
observed attenuation of charmonium production should be the same for
\J~as for \P, as it is indeed found to be (see Fig.\ 8 \cite{psi/psi'}).
Earlier attempts to explain charmonium suppression in $p-A$
interactions in terms of the absorption of physical \J~states
\cite{Huefner} had encountered difficulties precisely because of this
feature. The equal attenuation of \J~and \P is a natural consequence
of pre-resonance absorption; it can never be obtained for the
physical \J~and \P~states with their very different geometric sizes.

\bigskip

\begin{figure}[t]
\vspace*{2mm}
\centerline{\psfig{file=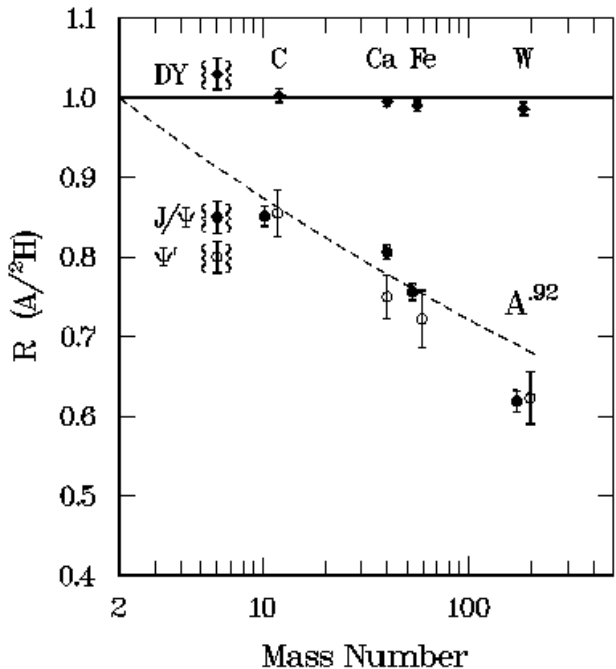,height= 90mm}\hspace{2cm}}
\caption{\J~production in $p-A$ vs. $p-p$ collisions \cite{E772}.}
\end{figure}

\medskip

The cross section for the dissociation of the $\C g$ state through
collisions in nuclear matter can be estimated theoretically \cite{KS6};
but it can also be determined directly from $p-A$ data \cite{KLNS}. A
$\C$ pair formed at point $z_0$ in the target nucleus has a survival
probability
\be
S^A_{\C g} = \exp -\left\{ \int_{z_0}^{\infty} dz~ \rho_A(z)~
\sigma_{\C g-N} \right\},
\label{9}
\ee
where the integration covers the path remaining from $z_0$ out of the
nucleus. The traversed medium of nucleus $A$ is parametrized through a
Wood-Saxon density distribution $\rho_A(z)$, and by comparing $S^A_{\C
g}$ with data for different targets $A$, the dissociation cross
section for $\C g-N$ interactions is found to be \cite{KLNS}
\be
\sigma_{\C g-N} = 7.3 \pm 0.6~{\rm mb}. \label{10}
\ee
In Fig.\ 9 it is seen that pre-resonance absorption with this cross
section agrees well with all presently available $p-A$ data on
\J~production.

\bigskip

\begin{figure}[p]
\vspace*{-0mm}
\centerline{\psfig{file=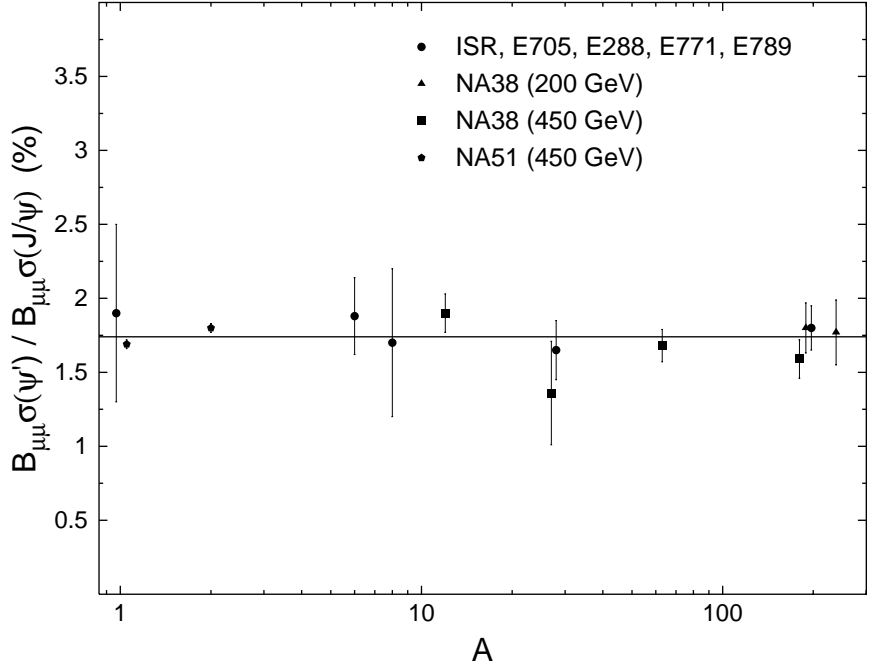,height= 90mm}\hspace{3cm}}
\caption{The relative $A$-dependence of \J~and \P~production.}
\end{figure}

\begin{figure}[p]
\vspace*{-0mm}
\centerline{\psfig{file=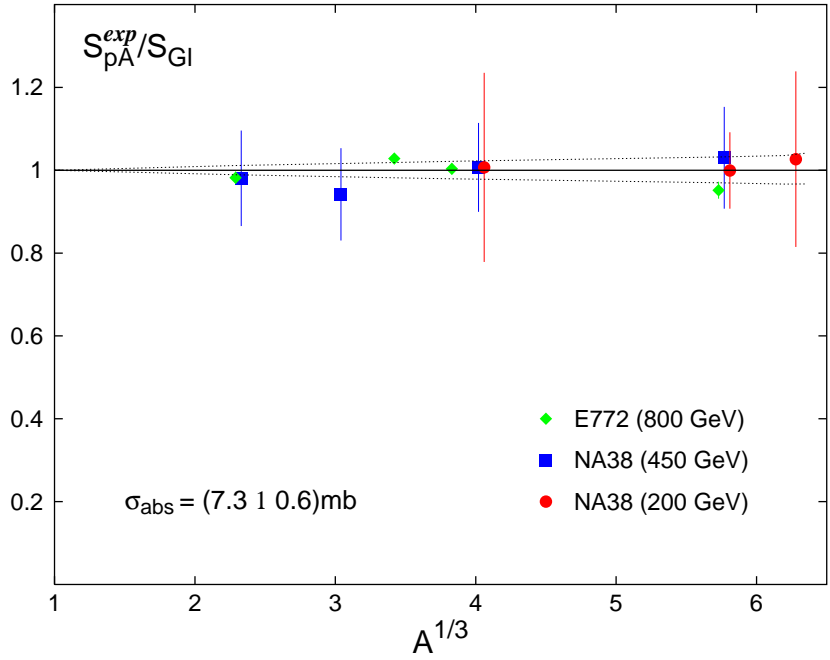,height= 90mm}}
\caption{\J~production in $p-A$ collisions, compared
to pre-resonance absorption in nuclear matter \cite{KLNS}.}
\end{figure}

\medskip

We now turn to nucleus-nucleus collisions; here there will certainly
also be pre-resonance absorption in nuclear matter. However, in addition
to the target and projectile nuclei, there could now be a substantial
amount of produced `secondary' medium (Fig.\ 10). We want to check if
there is such a medium and if yes, test its confinement status.

\bigskip

\begin{figure}[ht]
\vspace{2mm}
\centerline{\psfig{file=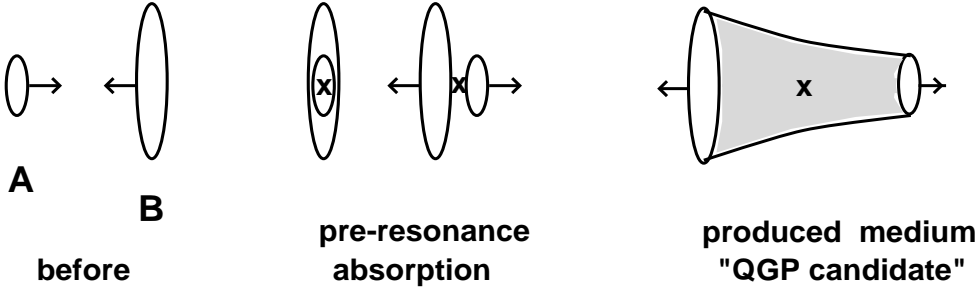,width=130mm}}
\vspace*{5mm}
\caption{Schematic view of \J~production in $A-B$ collisions.}
\end{figure}

\bigskip

The survival probability of a pre-resonance charmonium state in an $A-B$
collision at impact parameter $b$ is given by
\be
S^{AB}_{\C g}(b) = \exp \left\{- \int_{z_0^A}^{\infty} dz~ \rho_A(z)~
\sigma_{\C g-N}~ + ~\int_{z_0^B}^{\infty} dz~ \rho_B(z)~ \sigma_{\C g-N}
\right\},
\label{11}
\ee
in extension of Eq.\ (9). Here $z_0^A$ specifies the formation point of
the $\C g$ within nucleus $A$, $z_0^B$ its position in $B$. Since
experiments cannot directly measure the impact parameter $b$, we have
to specify how Eq.\ (10) can be applied to data.

\medskip

The Glauber formalism allows us to calculate the number
$N^{AB}_w(b)$ of participant (`wounded') nucleons for a given collision.
The number of secondary hadrons produced in association with the
observed \J~is
found to be proportional to $N^{AB}_w$. The transverse energy $E_T$
carried by the secondaries is measured experimentally, together with the
\J's. We thus have
\be
E_T(b) = q~N_w^{AB}(b); \label{12}
\ee
the proportionality constant $q$ has to be determined on the basis of
the given experimental acceptance. Once it is fixed, we have to check
that the collision geometry (the measured relation between $E_T$ and the
number of spectator nucleons, the measured $E_T$-distribution) are
correctly reproduced \cite{KLNS}. Once this is assured, we can check if
the \J~production in $O-Cu,~O-U$ and $S-U$, as measured by the NA38
experiment at CERN over the past ten years \cite{NA38}, shows anything
beyond the expected pre-resonance nuclear absorption with the cross
section $\sigma_{\C g-N} = 7.3 \pm 0.6$ mb determined from $p-A$
interactions.

\medskip

The answer is clearly negative, as seen in Fig.\ 11 for the integrated
cross sections and in Fig.\ 12 for the centrality ($E_T$) dependence of
$S-U$ collisions. All nuclear collisions measured by NA38 thus show only
what is now called `normal' \J~suppression, i.e., the pre-resonance
suppression already observed in $p-A$ interactions
\cite{KS6,Huefner,KLNS}. We thus have to
ask if the $A-B$ collisions studied by NA38 lead to any produced
secondary medium at all. The behaviour of \P~production (Fig.\ 13)
shows
that this is indeed the case \cite{NA38-psi'}: there is \P~suppression
beyond the expected pre-resonance nuclear absorption. There thus is a
secondary medium, and it can distinguish a \P~(which is suppressed)
from a \J~(which remains unaffected). In other words, this medium shows
no indication for colour deconfinement.

\bigskip

\begin{figure}[p]
\vspace*{-0mm}
\centerline{\psfig{file=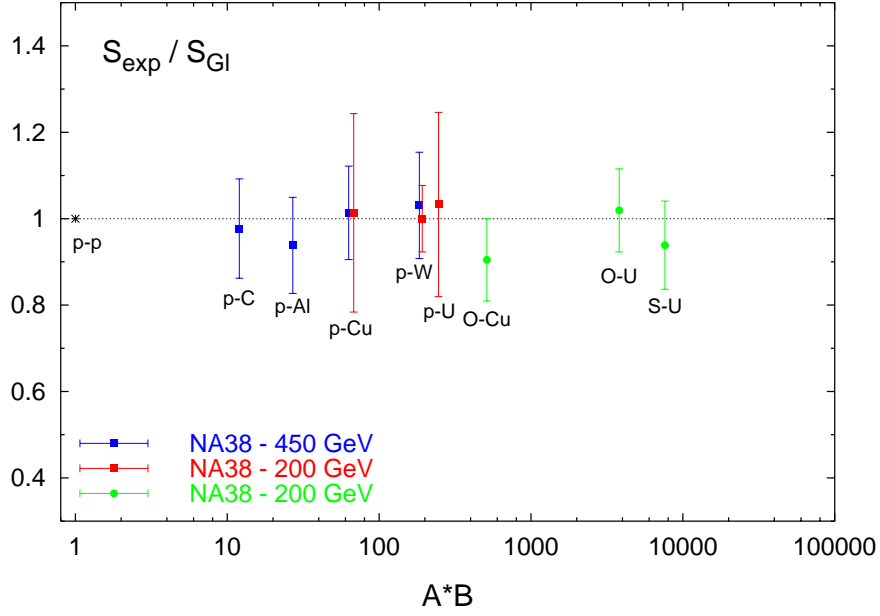,height= 90mm,angle= -90}}
\caption{\J~production in $A-B$ collisions, compared
to pre-resonance absorption in nuclear matter \cite{KLNS}.}
\end{figure}

\begin{figure}[p]
\vspace*{-0mm}
\centerline{\psfig{file=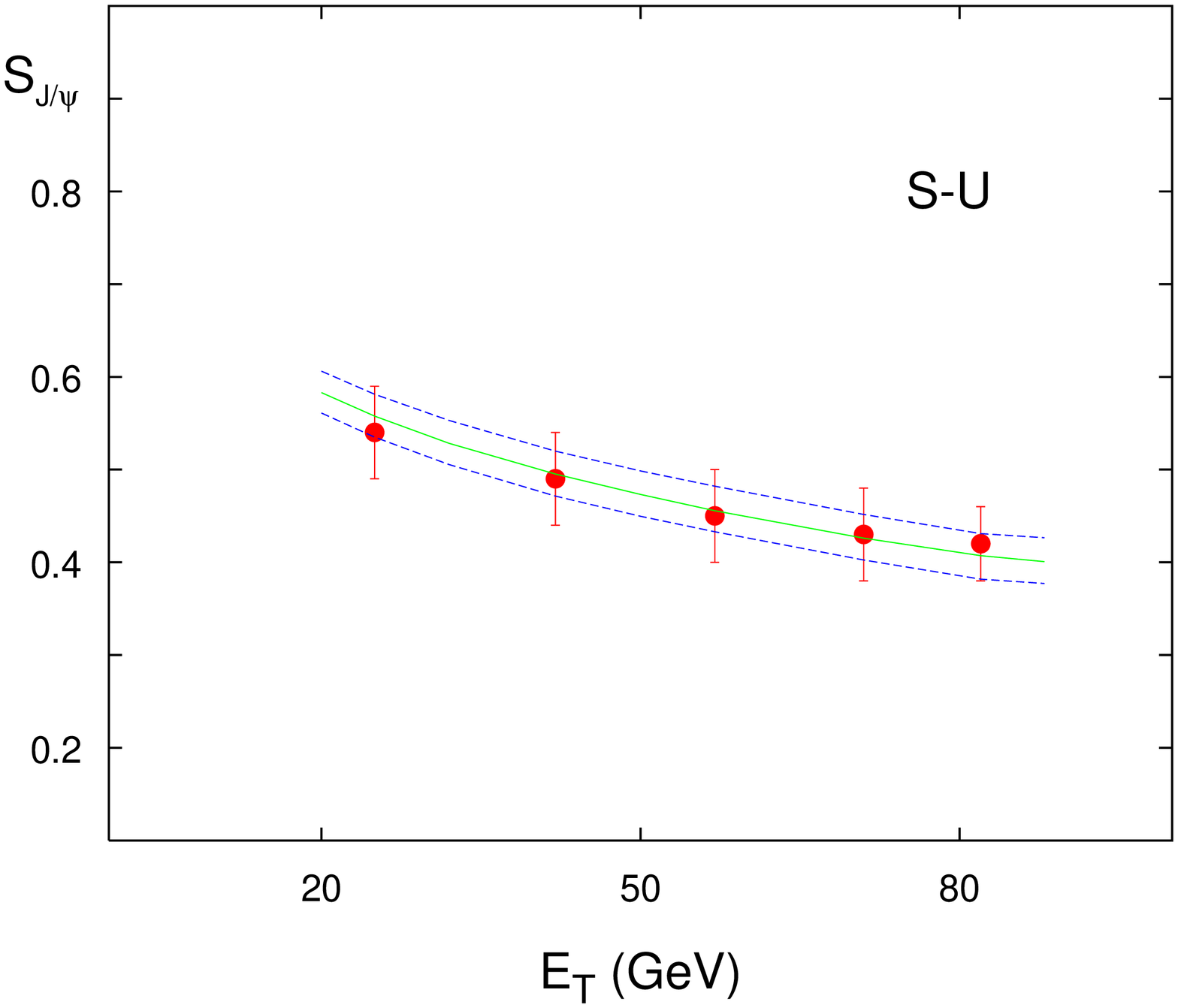,height= 90mm,angle= -90}, \hspace{1.5cm}}
\caption{The $E_T$-dependence of \J~production in $S-U$
collisions, compared to pre-resonance absorption in nuclear matter 
\cite{KLNS}.}
\end{figure}

\bigskip

\begin{figure}[t]
\vspace*{-0mm}
\centerline{\psfig{file=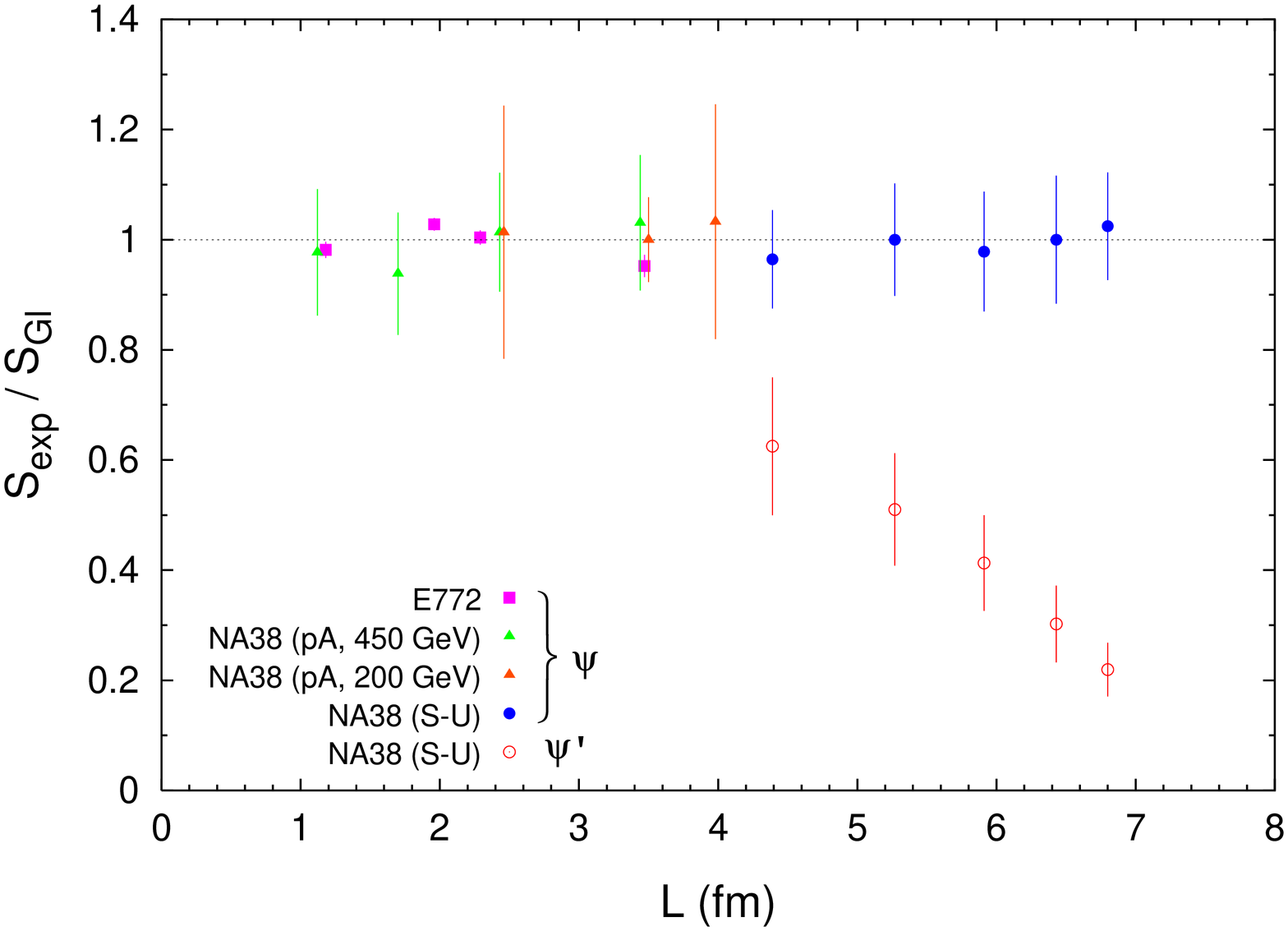,height= 100mm,angle= -90}}
\vspace*{-5mm}
\caption{\J~and \P~production in $A-B$ collisions \cite{NA38-psi'},
compared to pre-resonance absorption in nuclear matter \cite{KLNS}.}
\end{figure}

\noindent
{\bf 4.\ Anomalous \J~Suppression}

\medskip

In view of this state of affairs, the 30 \% additional (and hence
`anomalous') \J~suppression observed in $Pb-Pb$ collisions by the NA50
collaboration at CERN \cite{NA50} caused considerable excitement; the
results from the first (1995) run are shown in Fig's.\ 14 and 15. Is
this the first indication for the onset of colour deconfinement
\cite{Blaizot,DimaHeidel}?

\begin{figure}[p]
\vspace*{-0mm}
\centerline{\psfig{file=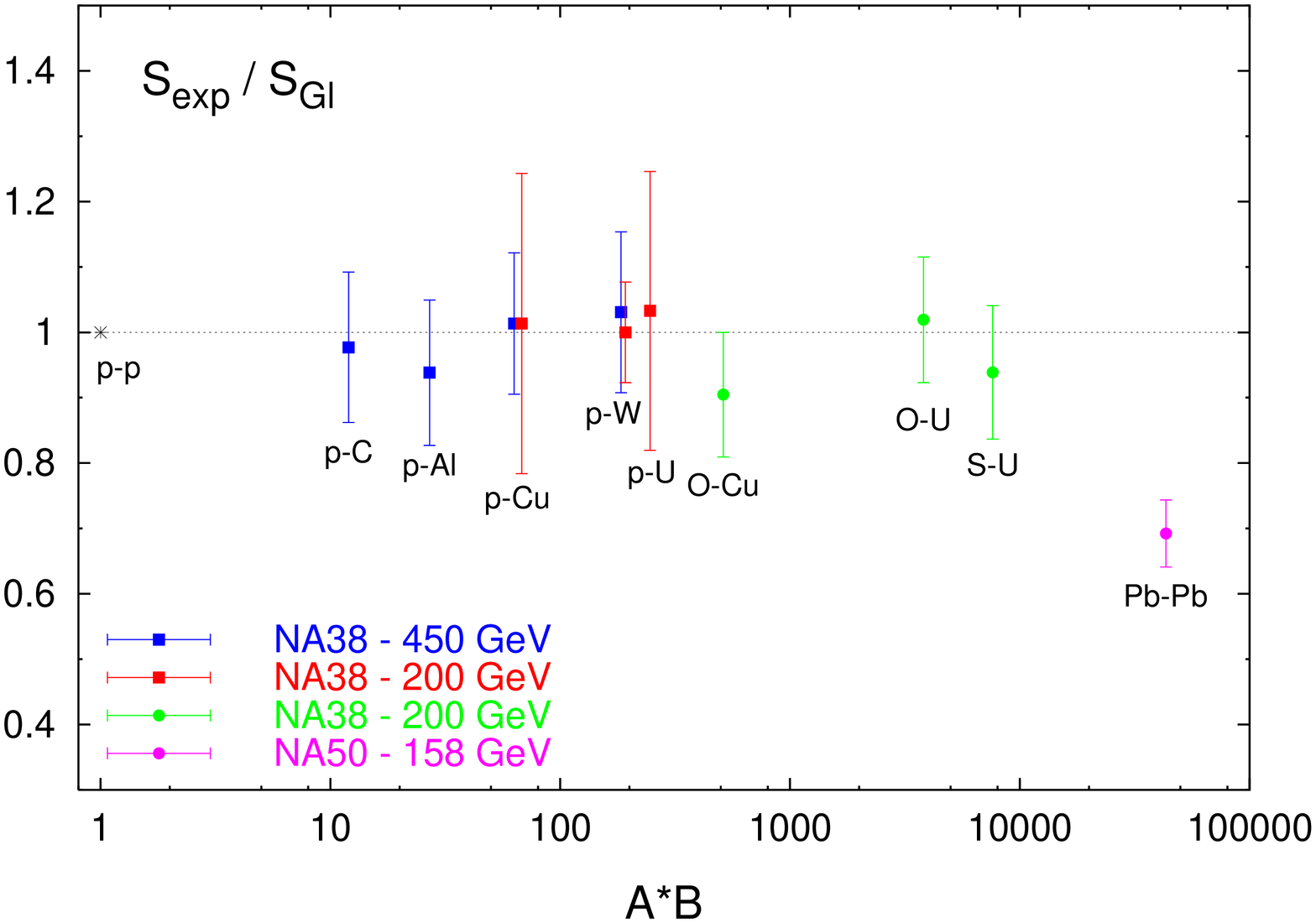,height= 90mm,angle= -90}}
\caption{\J~production in $A-B$ and $Pb-Pb$ collisions \cite{NA50},
compared to pre-resonance absorption in nuclear matter \cite{KLNS}.}
\end{figure}

\begin{figure}[p]
\vspace*{-0mm}
\centerline{\psfig{file=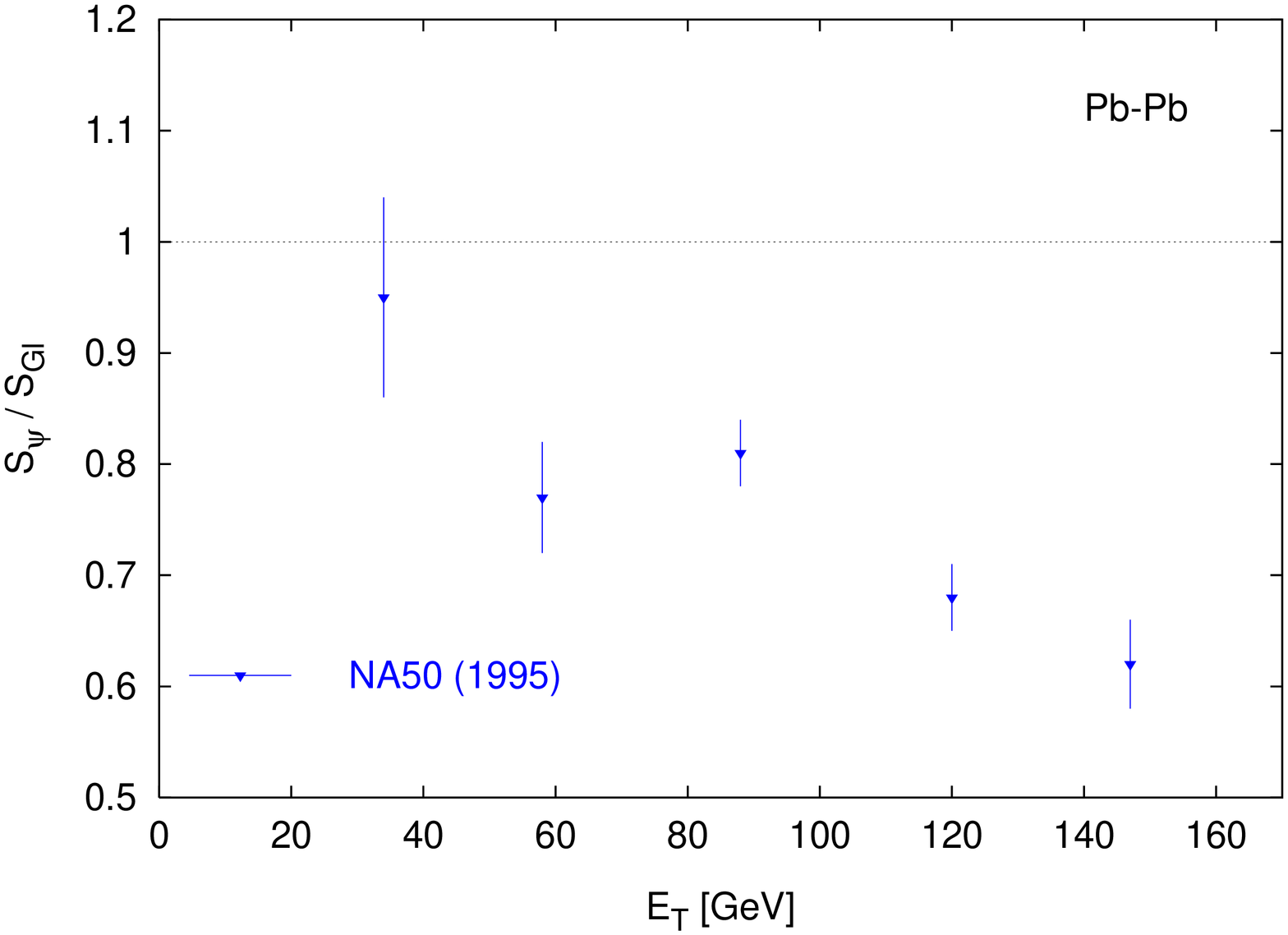,height= 90mm,angle= -90}}
\caption{The $E_T$-dependence of \J~production in $Pb-Pb$
collisions \cite{NA50}, compared to pre-resonance absorption in nuclear matter
\cite{KLNS}.}
\end{figure}

\medskip

To address this question, we have to check if it is possible to
interpret the observed effect in terms of dissociation in a confined
medium. The $Pb-Pb$ collision could produced a medium of secondary
hadrons, and the \J~could be broken up by interactions with these
hadronic comovers \footnote{~Such an approach is evidently not in accord
with the small hadron-\J~dissociation cross section obtained in Eq.\
(8); it thus implicitly assumes that the charm quark mass is not yet
large enough for the applicability of short-distance QCD calculations.}.
The main feature of such absorption is that it has no specific onset,
but is always operative; hence it would have to be present to some
extent also in $S-U$ collisions. Its functional form at fixed impact
parameter $b$ is \cite{Gavin,Capella}
\be
S_{\rm co}(b) = \exp - \left\{n_{\rm co}(b) \sigma_{co} \tau_0
\ln [n(\e)/n_f] \right\}, \label{13}
\ee
where $n(b)$ is the comover density, $\tau_0 \simeq 1$ fm the time
required to form the comover medium, and $n_f$ the comover density at
freeze-out. Eq.\ (13) always leads to a monotonic increase of
suppression with increasing centrality and comover density.

\medskip

\J~dissociation by colour deconfinement, in contrast, sets in at a
specific energy density of the deconfining medium (see Table 1 and
Fig.\ 16).
To illustrate the consequences, consider a central $A-A$ collision. The
energy density in the transverse plane will be highest at the center,
along the collision axis, and will decrease with increasing distance $r$
from the axis (Fig. 17). Once the central value passes the critical
deconfinement point, the \J's in the `hot' central region start melting,
while those in the cooler outer rim survive. If we take a hard sphere
nuclear distribution, this leads to
\be
S_{\psi}(\e) = \Theta(\e_{\psi} - \e) + \Theta(\e - \e_{\psi})
\left({\e_{\psi} \over \e} \right)^{9/4} \label{14}
\ee
for the survival probability of the \J~in a medium of central energy
density $\e=\e(r=0)$ \cite{GS1}. If we identify the deconfinement
threshold $\e_{\psi}$ with the energy density at $r=0$ in a central
$S-U$ collision and use more a realistic Wood-Saxon potential, then
this reproduces quite well the amount of suppression observed in
central $Pb-Pb$ collisions \cite{Blaizot}.

\bigskip

\begin{figure}[p]
\vspace*{-0mm}
\centerline{\psfig{file=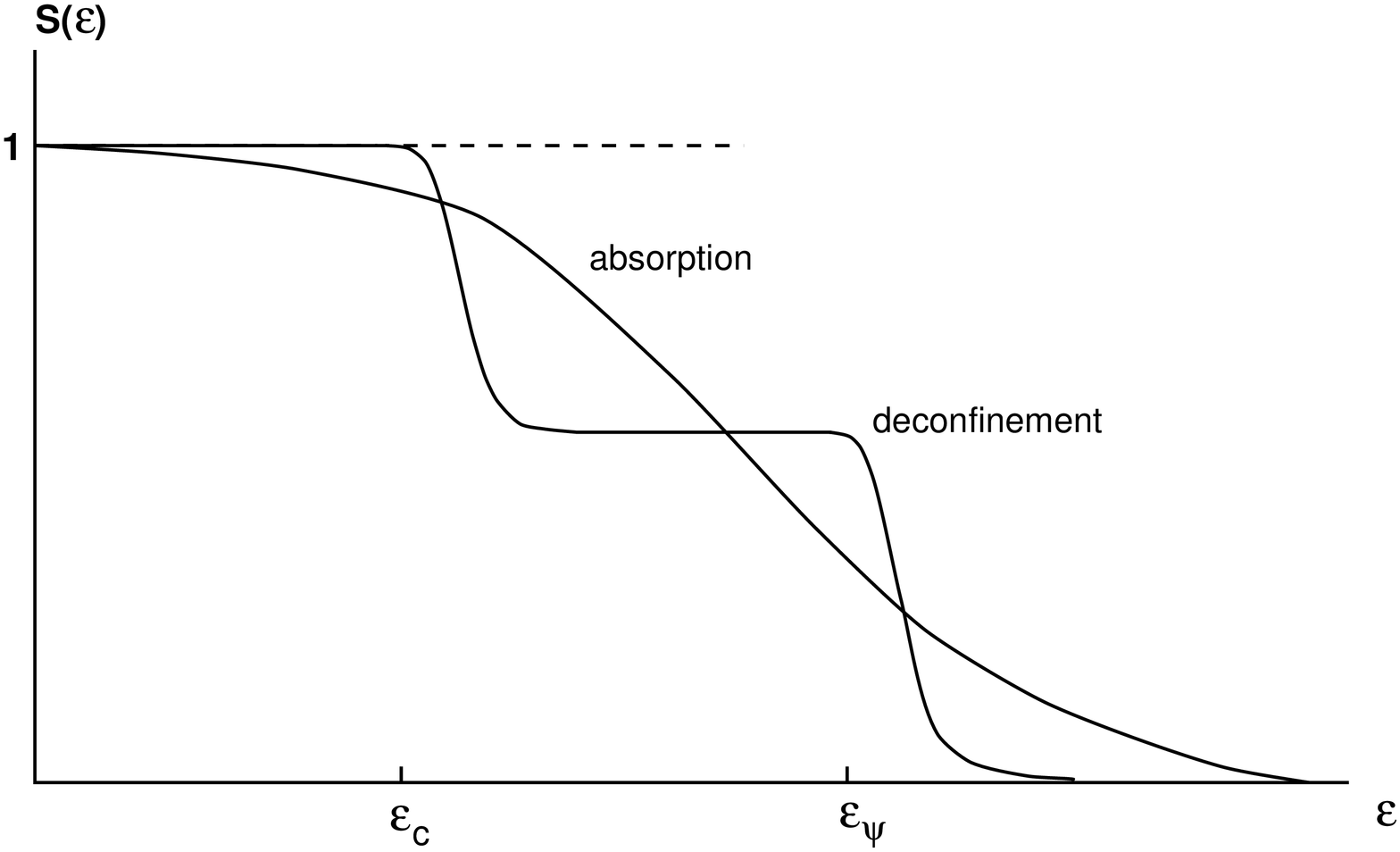,height= 70mm,angle= -90}}
\caption{\J~suppression by hadronic comover absorption compared to that
by colour deconfinement.}
\end{figure}

\begin{figure}[p]
\vspace{-20mm}
\centerline{\psfig{file=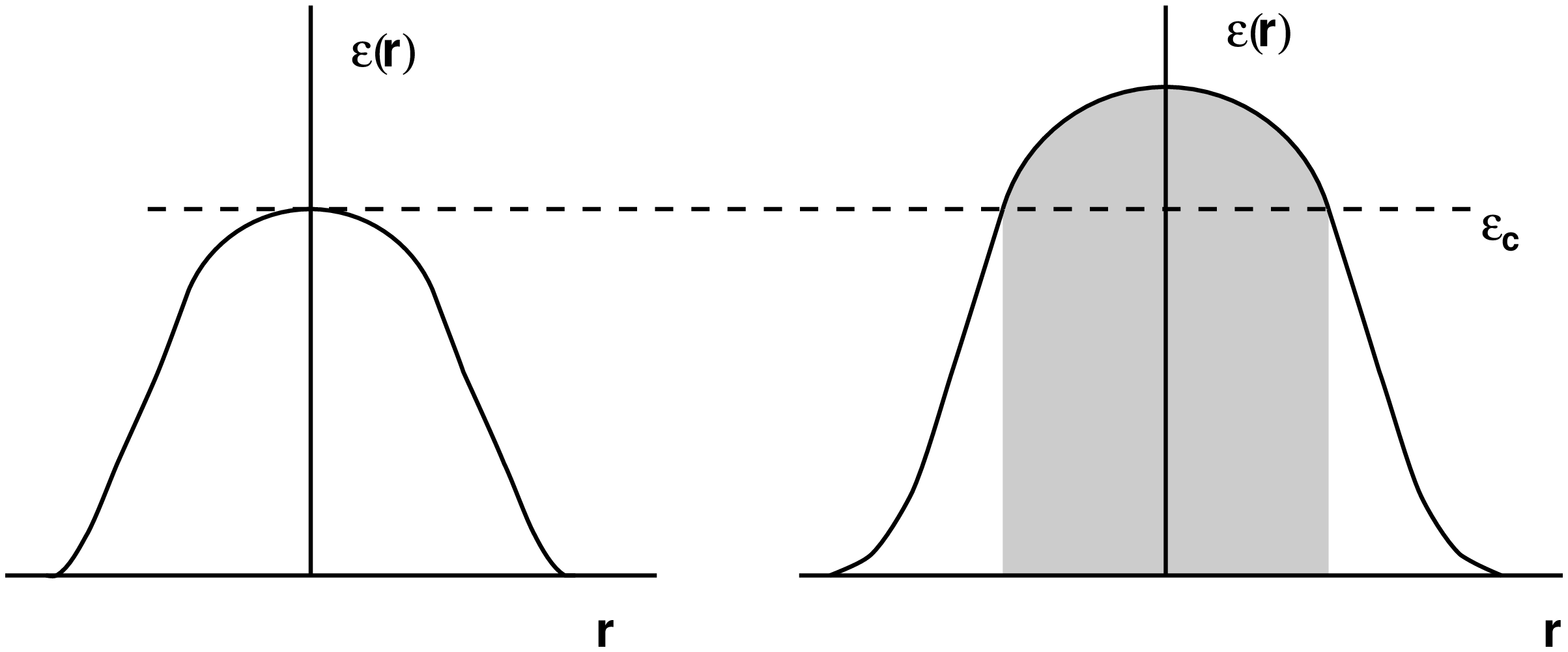,width= 120mm,angle= -90}}
\caption{Energy density profiles for central $A-B$ collisions
above and below the deconfinement threshold.}
\end{figure}

\medskip

To study the onset of deconfinement in more detail, we have to take into
account the fact that of the observed \J's, about 60 \% are directly
produced $1S$ states, while the remainder is due to $\x~ (\sim 30 \%)$
and $\p~(\sim 10 \%)$ decay. Since the latter melt earlier, this leads
to the two-step suppression pattern \cite{GS1,K&P} already mentioned
in section 2a:
once we pass the deconfinement threshold, first the \J's originating
from \X~and \P~decay melt, and then, for sufficiently high energy
density, also the directly produced $1S$ states start to disappear.

\medskip

The statistics of the data shown in Fig.\ 15 are clearly not sufficient
to test such details. However, the 1996 run of NA50 has provided more
than five times as many \J's (about 275 000), and the new but still
preliminary data look very interesting indeed \cite{newdata}. In Fig.\
18, we compare
it to a form based on a first order deconfinement transition with
critical bubble size formation \cite{KNS2}; we emphasize that this
Figure is only meant to illustrate the present preliminary experimental
pattern and the general theoretical behaviour due to a discontinuous
onset of deconfinement. A more detailed theoretical study will have to
wait for the final data analysis. We note here only that even for a
discontinuous deconfinement onset, the $E_T-b$ smearing needed to
relate theory to experiment will cause a considerable softening of the
discontinuity, as seen in Fig.\ 18. Nevertheless, if such a step-like
suppression pattern is confirmed by the final data, any hadronic comover
explanation is excluded already on a qualitative level.

\newpage

\begin{figure}[h]
\vspace*{-0mm}
\centerline{\psfig{file=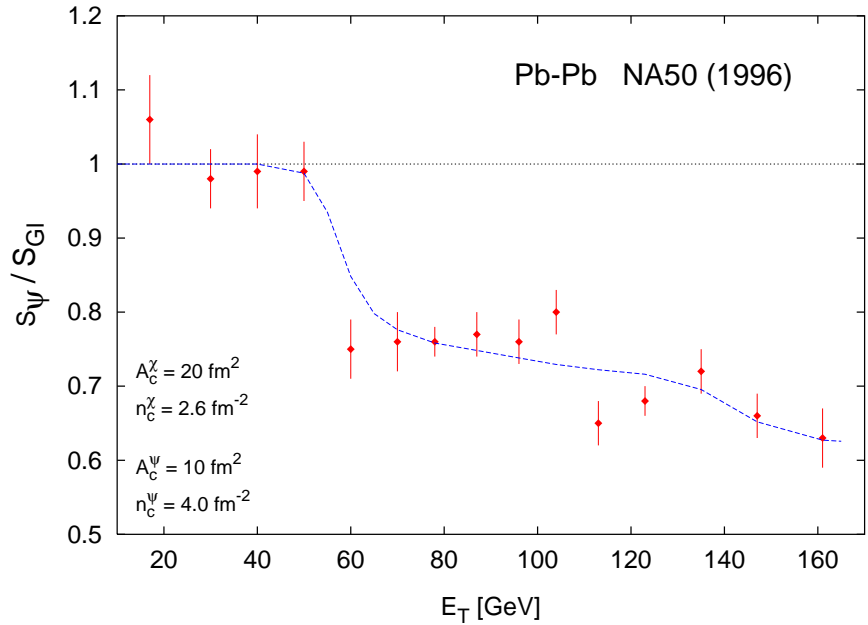,height= 90mm,angle= -90}}
\vspace*{-5mm}
\caption{Preliminary data on the $E_T$-dependence of
\J~production in $Pb-Pb$ collisions \cite{newdata}, compared to the pattern
from a first order deconfinement transition \cite{KNS2}.}
\end{figure}

\bigskip

\noindent
{\bf 5.\ Outlook and Summary}

\medskip

To establish that the observed anomalous \J~suppression is indeed due to
the onset of colour deconfinement, some further experimental studies
are clearly necessary.
\begin{itemize}
\item{First and foremost, it remains to be seen if the final analysis
of the 1996 $Pb-Pb$ run corroborates the step-like onset of
\J~suppression. To determine the crucial variable for a deconfinement
onset, one should then further study the effect for $A-A$ collisions
using different collision energies and different $A$.}
\item{Additional information will come from a study of the
dependence of the anomalous suppression on the transverse momentum of
the \J. Initial state multiple scattering of the gluons fusing to form
the $\C$ pair leads to a broadening of the $P_T$ spectra from nuclear
targets, compared to that in $p-p$ collisions. This `normal' $P_T$
broadening accounts for all observed \J~$P_T$-distributions up to
central $S-U$ collisions \cite{normal PT}. The melting process in
anomalous \J~suppression removes exactly those \J's (in the central
tube of the interaction region) which came from gluons with most initial
state scattering. This should lead to a characteristic turn-over of
$P_T$ broadening \cite{KNS1}, and it will be of great interest to see if
such an anomalous $P_T$ behaviour of \J~suppression is shown by the
data.}
\item{A further signal for a second step corresponding to the
onset of dissociation of directly produced \J's would be provided by an
increase in the ratio of \P~to \J~production.}
\item{Finally, it is obviously challenging to check if an onset of
colour deconfinement in the early stages of nuclear collisions, as are
probed by charmonia, will have repercussions also on other observables.}
\end{itemize}
\par\noindent
Some of these questions may well be answered within the next few
months, when the final analysis of the NA50 is presented. The search for
colour deconfinement thus appears indeed to have reached a rather
decisive stage.

\medskip

In summary: statistical QCD predicts colour deconfinement in media of a
sufficiently high density of constituents. Nuclear collisions are
expected to lead to such conditions, and quarkonium states provide a
probe to check the confinement/deconfinement status of the produced
medium.

\medskip

Up to central $S-U$ collisions, nucleus-nucleus data do not show any
\J~attenuation beyond `normal' pre-resonance absorption in the nuclear
matter of target and projectile. In contrast, $Pb-Pb$ collisions
appear to lead with increasing centrality to an abrupt onset of a
further `anomalous' suppression. Such a sudden change in behaviour has
no `conventional' explanation; if confirmed in the final data analysis,
it would indeed find a natural account only in terms of some form of
critical behaviour.

\bigskip
\noindent
{\bf Acknowledgements:}

\medskip

Much of what is presented here is based on joint work with D.\ Kharzeev
and M.\ Nardi; I thank them for this fruitful and enjoyable
collaboration. It is also a pleasure to thank M.\ Gonin, L.\
Kluberg and C.\ Louren{\c c}o for stimulating discussions of the
NA38/NA50 data.

\end{document}